\documentclass[final,2p,times,onecolumn]{article}

\usepackage{hyperref}
\usepackage{amssymb}
\usepackage[pdftex]{graphicx}
\usepackage{amsmath}
\usepackage{bm}
\usepackage{xcolor}
\usepackage[switch,mathlines]{lineno}
\usepackage[normalem]{ulem}

\usepackage{xcolor}

\title{Precision measurement of ocean tides using cosmic-ray muons}

\author{
Jiwon~Seo$^{1}$,
Jongseok~Chung$^{1}$,
Chang~Hyon~Ha$^{1}$,
TongZhou~Huang$^{1}$,\\
Jinyoung~Kim$^{1}$,
Sooa~Kim$^{1}$,
Hani~Kimku$^{1}$,
Byoung-cheol~Koh$^{1}$,\\
Minsu~Kwak$^{1}$,
Sihyun~Lee$^{1}$,
and Yujin~Lee$^{1}$\\[4pt]
\small $^{1}$Department of Physics, Chung-Ang University,
Seoul 06974, Republic of Korea\\
}

\date{}

\begin{document}

\maketitle

\begin{abstract}
  Cosmic-ray muons provide a natural probe of the integrated mass along their trajectories.
  Here we show that cosmic-ray muons can monitor ocean-tide-induced variations
  in seawater overburden within an underwater tunnel.
  The observed modulation closely tracks independent tide-gauge measurements,
  yielding an attenuation response coefficient of $k=43.2\pm1.0$~counts$\rm /h/m$ and
  resolving meter-scale variations in seawater overburden.
  A spatial muography scan validates the overburden model, confirming
  that the seawater contribution is essential to reproduce the observed attenuation.
  Unlike conventional tide gauges that measure sea level locally, muography measures the integrated overburden above the detector.
  These results demonstrate that muography can quantitatively resolve small, time-dependent variations in environmental overburden,
  providing a new approach to monitoring dynamic natural systems.
\end{abstract}

\section{Main}

Ocean tides arise from the gravitational interplay among the Earth, Moon, and Sun,
continuously redistributing enormous masses of seawater across the planet~\cite{10.1098/rsta.1966.0024}.
Such periodic mass redistribution provides a natural setting for time-resolved muon tomography (muography).
Muography has enabled imaging of static structures ranging from volcanoes to pyramids~\cite{Morishima:2017ghw,Tanaka:2023ami},
while time-resolved measurements have demonstrated sensitivity to evolving density distributions,
such as magma dynamics in active volcanoes~\cite{Tanaka:2014}.
However, resolving meter-scale variations in environmental overburden requires greater statistical precision.
Although recent studies have explored the possibility of tidal muography~\cite{Tanaka2021,Khaw:2025tor},
quantitative measurements of tidal variations remain challenging.
Here we present a precision measurement of ocean-tide modulation
using cosmic-ray muons in an underwater tunnel beneath the West Sea of Korea.

Cosmic-ray muons, produced by interactions of primary cosmic rays in the atmosphere,
can penetrate substantial thicknesses of dense matter~\cite{ParticleDataGroup:2024cfk}.
Their attenuation in matter enables quantitative measurements of
integrated density along the particle trajectory~\cite{10.1111/j.1365-246X.2010.04790.x}.
However, time resolved muography is complicated by intrinsic variations
in the atmospheric muon flux arising from temperature and pressure~\cite{COSINE-100:2020jml,Abdullrahman2025}.
Underwater tunnels provide a natural laboratory in which ocean tides periodically modify the seawater overburden
in a manner independently measured by tide gauges.

A stationary plastic scintillator detector operated continuously for 31.6 days,
revealing a periodic modulation of the muon flux correlated with seawater level.
We additionally performed a spatial muography scan by completing ten one-way traversals of the tunnel (five round trips in total)
with the same portable detector system,
following the approach developed for previous highway-tunnel measurements~\cite{Seo:2024ijp}.
This complementary measurement provides an independent validation of the overburden model used to interpret the tidal modulation.

\begin{figure*}[!htb]
  \begin{center}
      \includegraphics[width=0.99\textwidth]{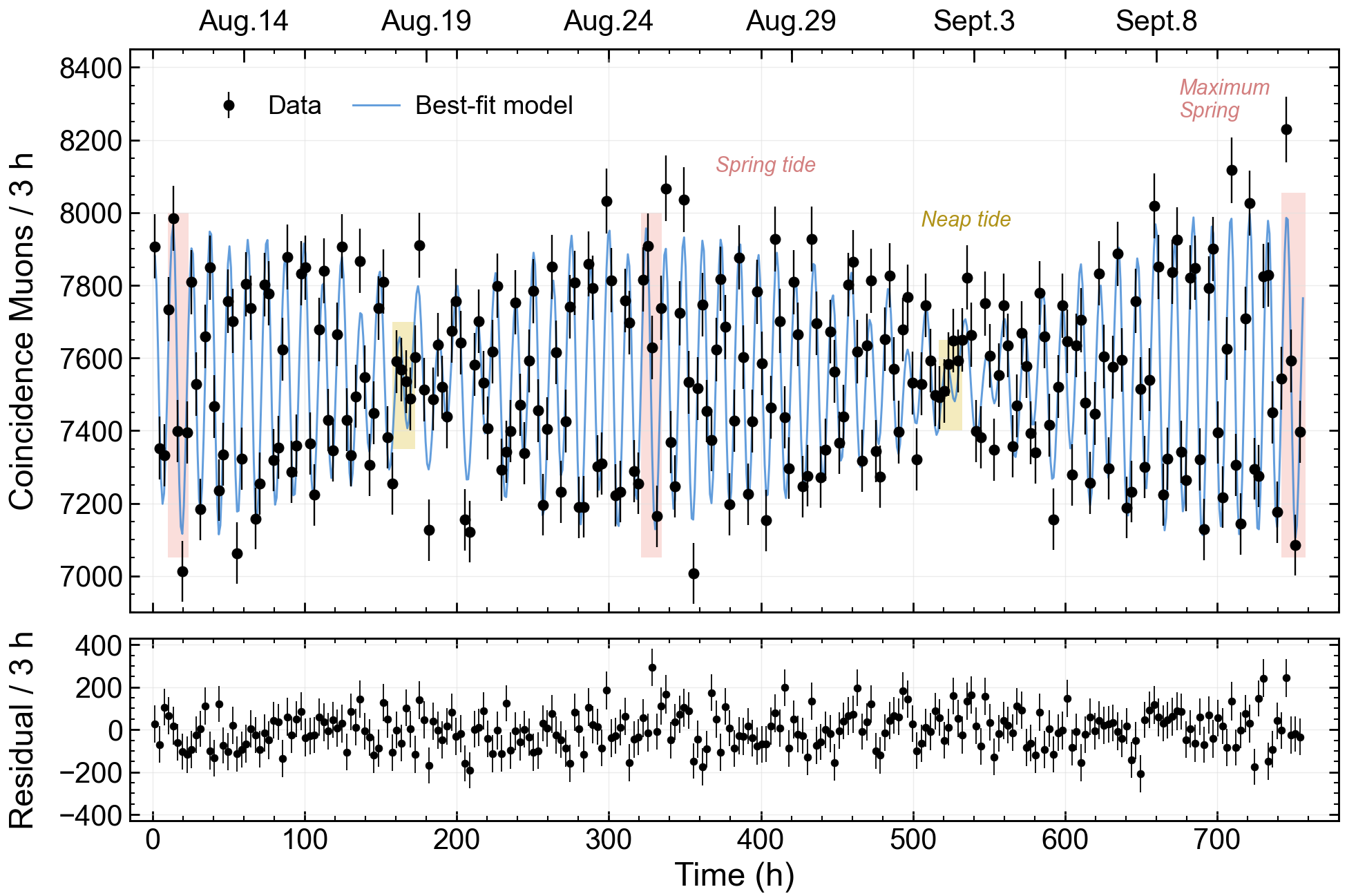}
  \end{center}
  \caption{
    Time variation of the muon rate measured with a stationary detector.
    The black points show the measured muon rate with statistical uncertainties while the blue line shows the best-fit model derived from the closest tide-gauge data,
    with the mean muon rate, overall modulation amplitude and phase treated as free parameters.
    The lower panel shows the residuals with respect to the best-fit model.
    The shaded regions indicate representative spring-tide and neap-tide periods.
    The measured muon rate is strongly anticorrelated with the seawater level ( correlation coefficient $r = -0.85$).
  }
  \label{hawl3}
\end{figure*}

\section{Dynamic Muography of Ocean Tides}
The measurement was performed in the Boryeong Undersea Tunnel, a 6.9 km long underwater tunnel~\cite{Korea-Tunnel}.
The large tidal variations, reaching approximately $\pm$4~m~\cite{TideData2}, periodically modulate the seawater overburden
and provide a favorable setting for studying time-dependent muon attenuation.
The detector consisted of two plastic scintillator panels,
each measuring $\rm 100\times50\times2~cm^3$, separated vertically by 42~cm~(Methods).
The detector was located at Station 5593~($36^\circ21^\prime09^{\prime\prime}\mathrm{N},
126^\circ27^\prime38^{\prime\prime}\mathrm{E}$), approximately 5.3~km from the Daecheon portal of the tunnel~\cite{Korea-Tunnel2}.
Near this location, the rock thickness above the tunnel is minimal at 26.0~m,
while the seawater thickness reaches its maximum of 33.8~m,
providing enhanced sensitivity to tidal water-level variations.
The detector operated continuously for 31.6 days between Aug. 11 and Sept. 12, 2025.
The full dataset contains approximately $1.9\times10^{6}$ detected coincidence muons.

Figure~\ref{hawl3} reveals a clear periodic modulation of the muon rate, with a characteristic timescale consistent with the local tidal cycle.
A likelihood fit to the observed data yields a mean muon rate of $R_0 = 2515\pm2$~counts/h,
an attenuation response coefficient of $k$ = 43.2$\pm$1.0~counts/h/m, and $\Delta t = -414\pm157$~sec~(Methods). 
The attenuation response implies that a 1~m rise in sea level reduces the detected muon rate by $1.72\pm0.04~\%$.
A dedicated Monte Carlo simulation in which the seawater level above the detector was varied
predicts a relative attenuation response of $1.89\pm0.03({\rm stat.}) {}^{+0.02}_{-0.03}({\rm syst.})$~\%/m,
where the systematic uncertainty reflects the variation of the assumed rock density from its nominal value of $2.65~\mathrm{g/cm^3}$,
consistent with the measured response~(Methods).
The fitted phase offset indicates that the tidal variation at the detector site leads the Boryeong tide gauge
by 6.9$\pm$2.6~min, corresponding to a 2.6~$\sigma$ deviation from zero.
The model provides a good description of the data, with $\chi^2/\mathrm{ndf}$= 768.1/755.
The residuals have a root-mean-square of 2.0~\% and
exhibit no statistically significant structure throughout the full observation period.

Fourier analysis further identifies the expected semidiurnal ($\rm 12.3\pm0.2$~hours) and spring–neap ($16\pm8$~days) components,
consistent with the principal lunar tide and the 14.8~days modulation
within the frequency resolution of the measurement~\cite{Renault2022,KVALE20065}~(Methods).
No significant diurnal atmospheric component is observed~(Methods).
Together, the time-domain, frequency-domain, and correlation analyses identify
changes in seawater overburden as the origin of the observed rate variation.

\section{Static Muography of Tunnel Overburden}
\begin{figure}[!htb]
  \begin{center}
    \includegraphics[width=0.48\textwidth]{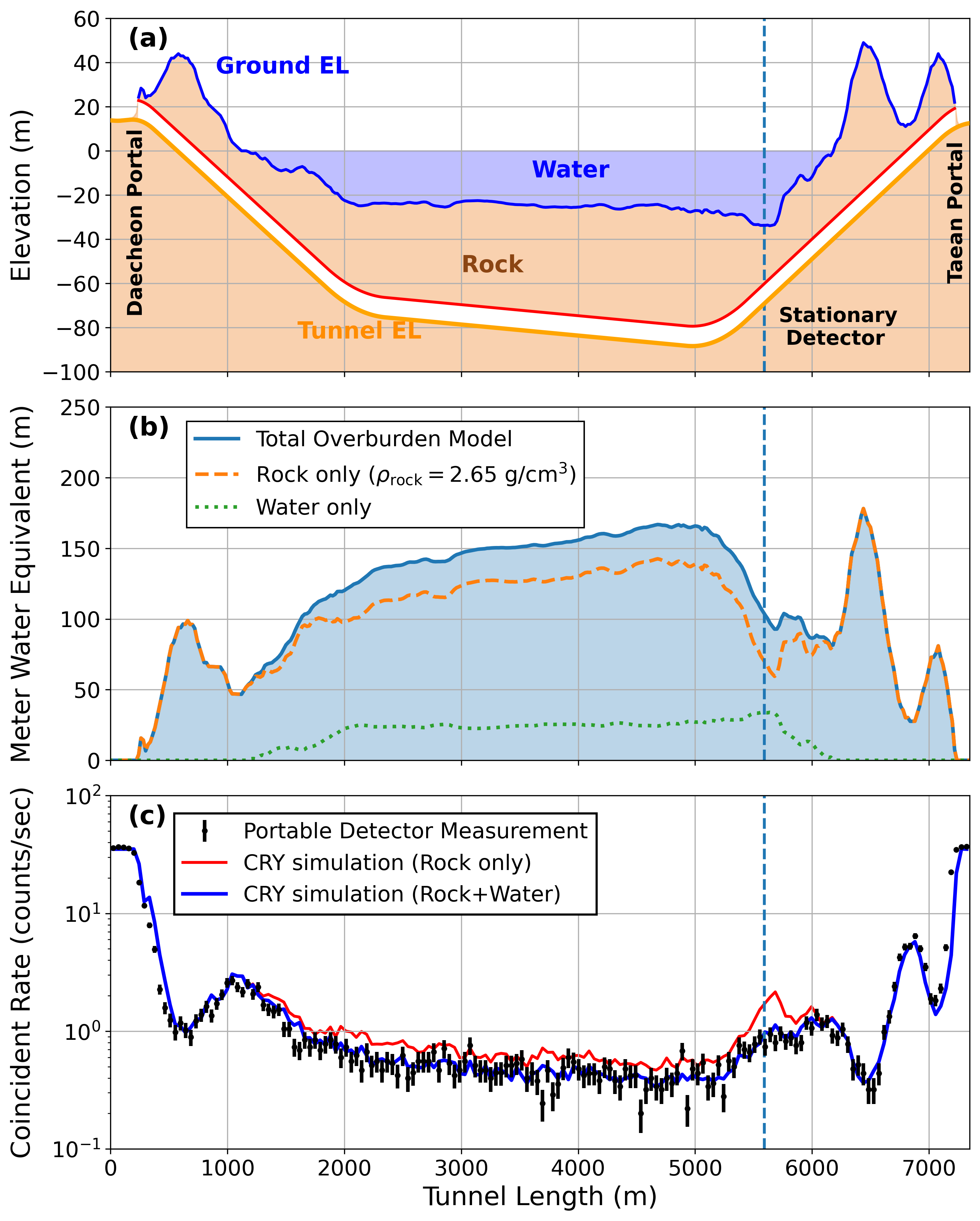}
  \end{center}
  \caption{Spatial muography scan of the underwater tunnel.
    (a) Elevation profile of the tunnel and ground along the tunnel axis.
    The shaded regions indicate the rock and seawater overburden above the tunnel.
    The vertical dashed line marks the stationary detector location (Station 5593),
    where the tide measurement was performed.
    (b) Estimated overburden expressed in meter-water-equivalent (m.w.e.).
    The total overburden model is separated into contributions from rock and seawater.
    An effective rock density of $\rho = 2.65~\mathrm{g/cm^3}$ is assumed~\cite{ART002186085}.
    (c) Measured coincidence muon rate as a function of tunnel position compared
    with the Monte Carlo expectation based on the overburden model.
    The attenuation of the muon flux measurements follows the variation in the tunnel overburden,
    with the lowest rates observed near the deepest tunnel section.
    The red curve shows the rock-only model, whereas the blue curve includes both rock and seawater overburden.
  }
  \label{calib}
\end{figure}

The rock and seawater overburden used to model the muon flux at the stationary detector was
independently validated using a spatial tunnel scan with a portable detector, as shown in Fig.~\ref{calib}.
The detector was moved along the full tunnel length and the measured coincidence rate
was compared with the CRY-based overburden simulation~\cite{Hagmann:2007ziw}~(Methods).
The measured attenuation profile reproduces the major features of the simulation,
including the large attenuation beneath the central underwater section and the recovery toward both tunnel portals.

The rock-only model fails to reproduce the measured spatial attenuation profile.
Relative to the nominal rock-plus-seawater overburden model, it gives a worse likelihood by $\Delta (-\ln L)$=230,
confirming that the seawater contribution is required to describe the data.

The spatial scan is, therefore, consistent with the expected contribution of the seawater to the observed muon attenuation.

\section{Discussion}

Time-resolved muography has previously been used to probe evolving density distributions,
particularly in volcanic systems~\cite{Tanaka:2014}.
Here, we extend this approach to environmental overburden variations on a substantially smaller scale,
resolved through their modulation of the underground muon rate.
The measured attenuation response corresponds to meter-scale variations in seawater overburden,
demonstrating quantitative sensitivity to percent-level changes in muon transmission.

The fitted phase offset is consistent with the regional tidal phase progression observed among nearby tide-gauge stations~(Methods).
If the 6.6~km separation from the Boryeong tide gauge is adopted as an effective propagation distance,
the measured delay corresponds to an apparent tidal phase speed of approximately $16\pm6$~m/s,
consistent with shallow-water tidal propagation in the region~\cite{PughWoodworth2014}.

Atmospheric and frequency-domain tests further support the interpretation of the observed rate variation
as a seawater-overburden effect~(Methods).

During the largest tidal excursion in the dataset, the measured coincidence rate increased by nearly 9\% above the average.
Using the measured attenuation response coefficient, this corresponds to a temporary reduction of approximately 5~m in seawater overburden
relative to the mean level. This illustrates the detector's sensitivity to transient changes in integrated seawater overburden.

Although this study uses only the integrated coincidence rate,
the same experimental concept can be extended to directionally resolved measurements.
Angular reconstruction would provide direction-dependent column densities,
while repeated spatial measurements could further constrain the spatial evolution of the overburden.
Such measurements could distinguish changes in total overburden from changes in its spatial distribution and
provide information beyond that available from a local sea-level measurement.

The present results demonstrate that time-dependent variations in environmental overburden
can be quantitatively resolved through cosmic-ray muon attenuation.
Ocean tides provide a particularly well-constrained example, in which an independently measured environmental variation
produces a predictable modulation of the underground muon flux.
This sensitivity to time-dependent column density complements established applications of muography
to larger-scale evolving structures and suggests opportunities for monitoring other natural systems
with variable overburden.

\section{Methods}

\subsection{Detector and Data Acquisition} 

The muon detector consists of two rectangular plastic scintillator (PS) panels (Eljen Technology,
EJ-200)~\cite{ejps} with dimensions of $1000 \times 500 \times 20~\mathrm{mm}^3$ each.
Each panel is instrumented with 40 silicon photomultipliers (SiPMs) arranged in a $10 \times 4$
grid with an approximately 100 mm pitch. The SiPMs (Hamamatsu S13360-6075PE) have an active
area of $6.0 \times 6.0~\mathrm{mm}^2$ and a gain of approximately $4.0 \times 10^6$ ~\cite{sipm}.

\begin{figure}[!htb]
  \begin{center}
    \includegraphics[width=0.48\textwidth]{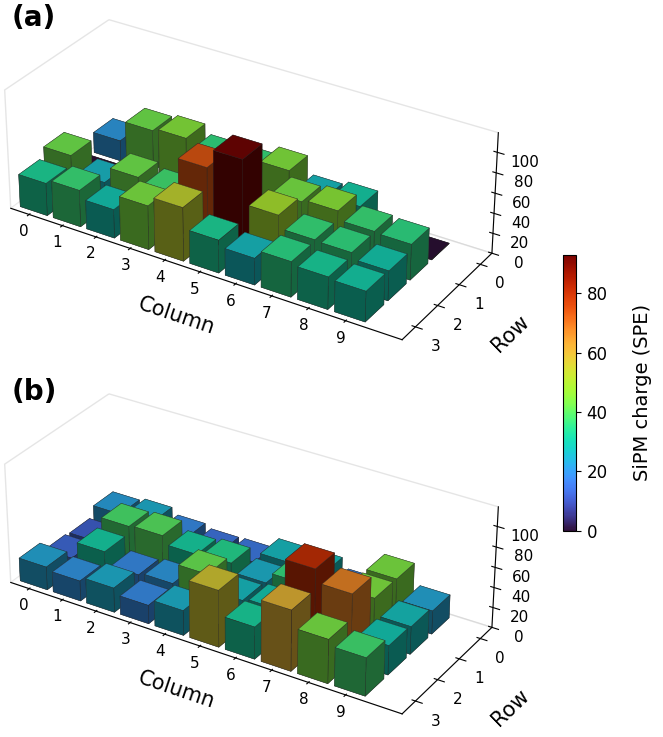}
  \end{center}
  \caption{Representative cosmic-ray muon event recorded by the detector.
    Gain-corrected SPE distributions are shown for the (a) upper and (b) lower plastic-scintillator panels.
    The spatially localized signals in both panels illustrate a coincidence event produced by a muon traversing the detector.
  }
  \label{event}
\end{figure}
Figure~\ref{event} shows a representative coincidence-muon event recorded by the detector.
The spatial distributions of the gain-corrected SiPM signals in the upper and lower scintillator panels
provide a two-dimensional representation of the light response induced by the traversing particle.

A channel hit is defined when the integrated signal charge exceeds 200~analog-to-digital converter counts within a 100~ns sliding window.
An event trigger is generated when 8 or more hits among the 80 active channels occur within a 1000~ns coincidence window.
For each trigger, a 4~$\mu$s readout window is recorded with the trigger time aligned near 1.2~$\mu$s.
The trigger energy threshold is about 1~MeV.

The trigger threshold was optimized to maintain a stable trigger rate below 2 kHz while preserving 99\% muon detection efficiency.
The expected coincident cosmic-ray muon rate at surface elevation is approximately 36~counts/s,
with an average deposited energy of approximately 4~MeV per scintillator panel~\cite{ParticleDataGroup:2024cfk}.

A calibration scan of the tunnel overburden was performed on June~30, 2025 using a movable detector
configuration. The stationary tide measurement was performed between Aug.~11 and Sept.~12, 2025.

Gain corrections associated with temperature variations of $(28\pm1)~^\circ\mathrm{C}$ were applied
offline using the measured single-photoelectron response of each SiPM channel.
The Compton edge of the 2614 keV gamma ray from $^{208}$Tl was used to monitor the temperature dependence of the detector response~\cite{Be2004,Knoll2010}. Muon candidate events were required to have per-panel energy larger than 2.6~MeV in both panels.
The coincidence muon rate measured at Station 5593 was 0.698~counts/s in the long-term tidal measurement,
with an estimated gamma-ray contamination below 0.1\%.
The short-term moving scan measured the consistent rate of $0.74\pm0.12$~counts/s at the same location.

\subsection{Tide Model and Statistical Analysis}

Tide data were obtained from the Boryeong tide gauge station, located approximately 6.6 km from
the detector site and operated by the Korea Hydrographic and Oceanographic Agency (KHOA).
The Boryeong station is the closest tide gauge to the Boryeong Undersea Tunnel.

\begin{figure}[!htb]
  \begin{center}
    \includegraphics[width=0.48\textwidth]{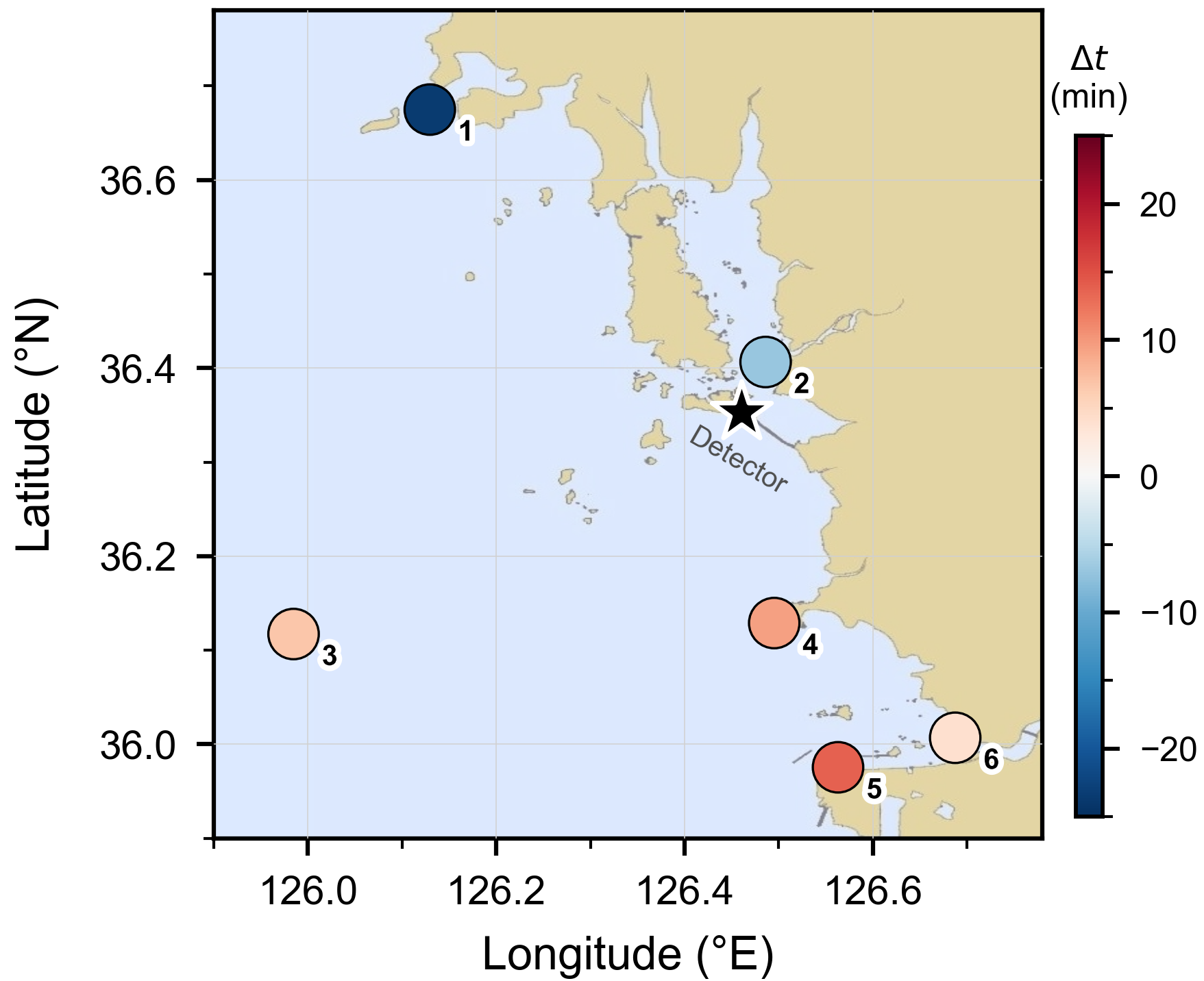}
  \end{center}
  \caption{
    Fitted tidal time shifts for nearby tide-gauge stations.
    The black star marks the location of the muon detector in the Boryeong Undersea Tunnel, while the yellow and blue regions denote land and sea respectively.
    The tidal time shift was determined by fitting the muon data to records from six nearby tide-gauge stations~\cite{TideData},
    with the marker colour indicating the fitted time shift for each station.
    The uncertainties in the fitted time shifts were approximately 2.6 min for all six stations.
    Negative values indicate that the tidal variation at the gauge station occurs later than at the detector location,
    whereas positive values indicate that it occurs earlier.
    The fitted time shift exhibits a strong negative correlation with latitude~($r=-0.96$).
  }
  \label{3dfigure}
\end{figure}

Several additional nearby tide stations were also examined.
The relative tidal phase differences among these stations were found to be consistent with model expectation
as shown in Fig.~\ref{3dfigure}.

The Boryeong tide-gauge data provide seawater-level measurements at 1~minute intervals.
A spline interpolation of the measured tide data was used to construct the time-dependent tidal template function $h(t)$.

The data were binned in 1~hour intervals. For each time bin $i$,
the observed number of coincidence muons is denoted by $n_i$,
and the expected number from the gauge-based seawater-height model
is denoted by $\mu_i$. The expectation was modeled as
\[
\mu_i =R_0 - k\,h(t_i-\Delta t),
\]
where $R_0$ is the mean muon rate, $k$ is the attenuation response
coefficient, and $\Delta t$ is the phase offset.

The fit was performed using a Poisson likelihood~\cite{BAKER1984437}. We used the
saturated Poisson likelihood ratio,
\[
-2\ln L =2\sum_i\left[\mu_i - n_i+n_i \ln\left(\frac{n_i}{\mu_i}\right)\right],
\]
where the saturated model is defined by $\mu_i^{\rm sat}=n_i$ in each
bin. For bins with $n_i=0$, the logarithmic term is taken to be zero.
The best-fit parameters were obtained by minimizing $-2\ln L$.
Profile-likelihood scans were used to estimate confidence intervals.
For one fitted parameter, the $1\sigma$ interval corresponds to $\Delta(-2\ln L)=1$~\cite{ParticleDataGroup:2024cfk}.

\subsection{Atmospheric and Environmental Systematics}

Atmospheric muon fluxes can vary with atmospheric pressure and temperature,
potentially introducing systematic effects into the observed modulation.
To quantify these effects, additional terms associated with atmospheric pressure and temperature
were introduced into the likelihood model,
\begin{equation}
  R(t) = R_0 - k\,h(t_i-\Delta t)- \alpha_x X(t_i),  
\end{equation}
where $X(t_i)$ represents either the atmospheric pressure $P$ or atmospheric temperature $T$
at an altitude of 13~km~\cite{ERA5PressureLevels,Hersbach2020}, and $\alpha_x$ denotes the corresponding response coefficient.
The fitted response coefficients were found to be consistent with zero and we set
90\% upper limit at $\alpha_P$ = $0.6$~counts$\rm/h/hPa$ and $\alpha_T$ = $0.6$~counts$\rm/h/^{\circ}C$.
The best-fit tidal response coefficient $k$ remained stable within uncertainty after inclusion of the additional atmospheric terms.

\begin{figure}[!htb]
  \begin{center}
    \includegraphics[width=0.48\textwidth]{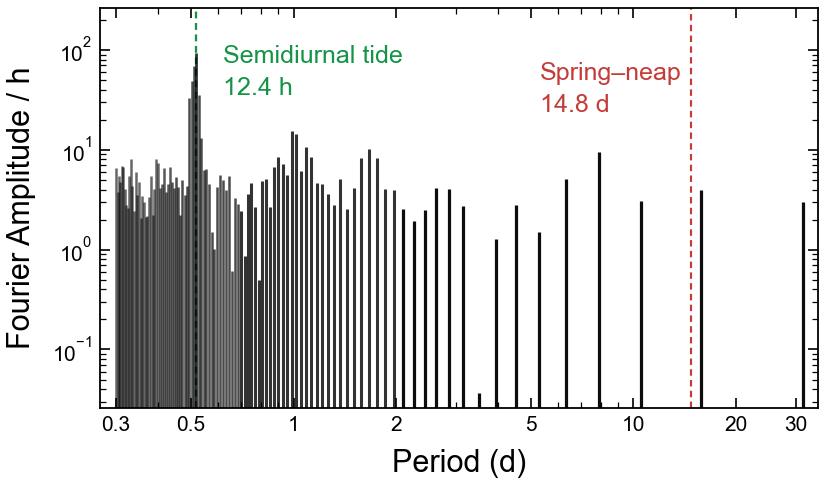}
  \end{center}
  \caption{Fourier amplitude spectrum of the hourly muon rate.
    The spectrum was computed from the hourly muon rate after linear detrending and application of a Hann window.
    The green and red vertical lines indicate the semidiurnal tidal period (12.4 h) and the spring–neap cycle (14.8 d), respectively.
    The dominant peak coincides with the semidiurnal tidal period.
    Because the 14.8-day spring–neap period falls between adjacent discrete Fourier modes for the 31.6-day observation,
    its power is represented by the nearest Fourier component.
  }
  \label{fft}
\end{figure}

The FFT spectrum (Fig.~\ref{fft})~\cite{d3ea2d52-5ab2-3128-8b80-efb85267295d,Harris1978} exhibits a dominant semidiurnal peak
at 12.4~hours together with the characteristic spring-neap modulation at 14.8~days.
These independent tests support the interpretation that the observed modulation is dominated by variations in seawater overburden.

\subsection{Overburden Modeling and Cosmic-Ray Simulation}

The spatial overburden above the Boryeong Undersea Tunnel was modeled using the tunnel design
elevation profile together with the local ground and sea floor geometry.
The total overburden was expressed in meter-water-equivalent (m.w.e.) units
by combining the rock and seawater contributions above the tunnel.

For the geological overburden, an effective rock density of $2.65~\mathrm{g/cm^3}$
was assumed to account for the average composition of the rock layer above the tunnel~\cite{ART002186085_2}.
The resulting total overburden reaches approximately 170 m.w.e. near the deepest central section of the tunnel.

Monte Carlo simulations of cosmic-ray muons were performed using the Cosmic-Ray Shower Library (CRY) generator
to model both the spatial attenuation along the tunnel and the tidal response at Station 5593.

The measured coincidence muon rate along the tunnel axis was compared
with the simulated expectation based on the overburden model.
The calibration scan data were found to be consistent with the
measured stationary coincidence rate at Station 5593 within uncertainty.

The tide analysis presented in this work primarily relies on the integrated muon attenuation
at a fixed detector location rather than on high-resolution angular reconstruction.
Consequently, the limited angular resolution of the detector does not significantly
affect the interpretation of the observed tidal modulation signal.

\begin{figure}[!htb]
  \begin{center}
    \includegraphics[width=0.48\textwidth]{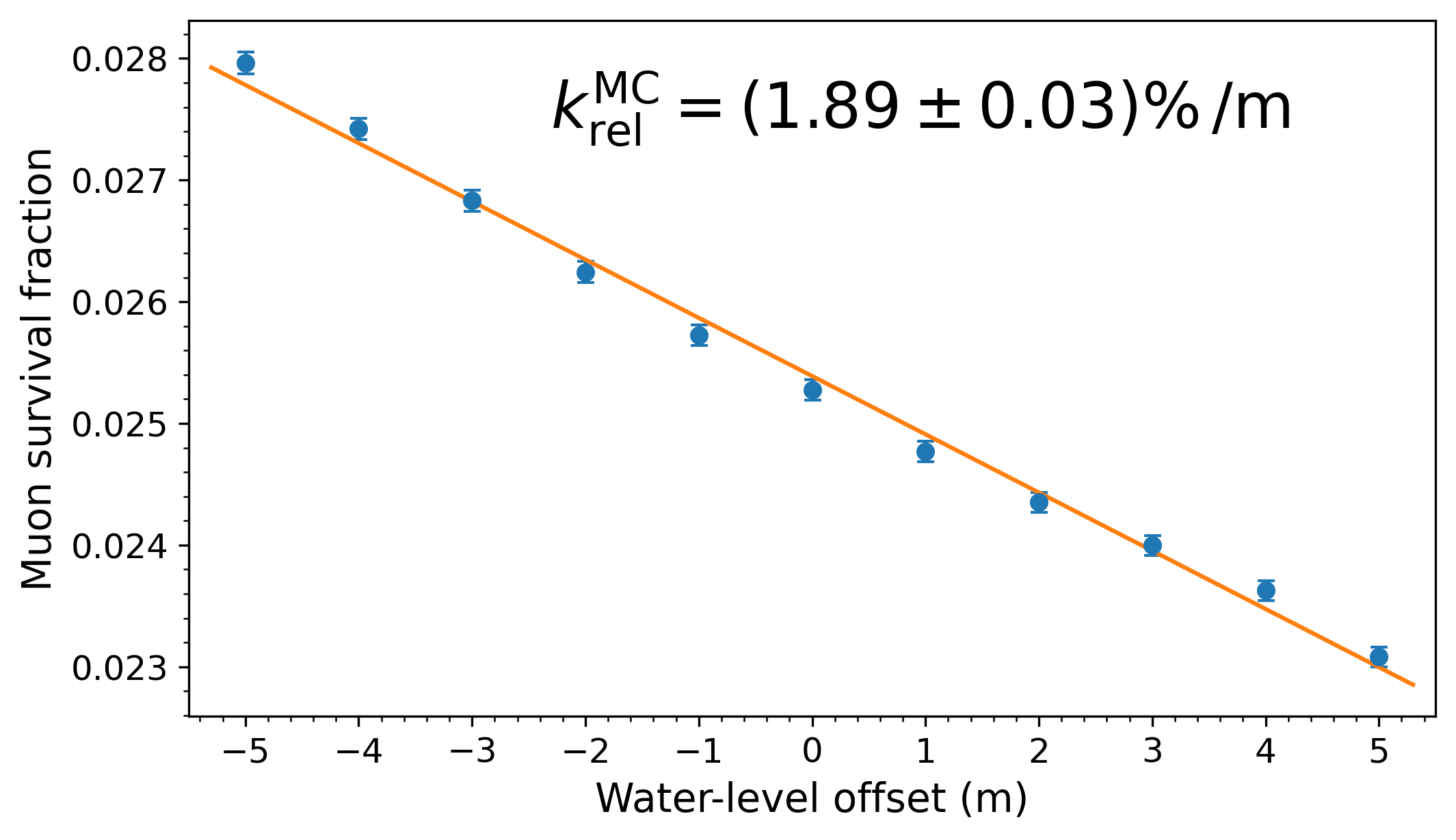}
  \end{center}
  \caption{Simulated dependence of the muon survival fraction on seawater level at the stationary detector location.
    Independent CRY samples were generated for 11 water-level offsets from $-5$ to $+5$~m.
    Points show the simulated survival fractions with statistical uncertainties, and the line shows a linear fit.
    The fitted relative attenuation response is $k^{MC}_{rel} = 1.89\pm0.03_{\rm stat.}$\%/m.
  }
  \label{crysim}
\end{figure}

A dedicated simulation was additionally performed at the stationary detector location
to quantify the expected response to changes in seawater level.
The local rock geometry was held fixed, while the seawater level was varied relative to its reference value.
Independent CRY muon samples were generated for each water-level offset and
propagated through the corresponding rock and seawater overburden.
Events satisfying the detector coincidence condition were then selected.
The simulated muon survival fraction was fitted as a function of the water-level offset,
yielding a relative attenuation response of $k^{MC}_{rel} = 1.89\pm0.03({\rm stat.}) {}^{+0.02}_{-0.03}({\rm syst.})$~\%/m,
which can be directly compared with the measured relative response $k/R_0$ (Fig.~\ref{crysim}).
The systematic uncertainty associated with the overburden model was evaluated
by varying the assumed rock density from the nominal value of $2.65~\mathrm{g/cm^3}$ to 2.60 and $2.70~\mathrm{g/cm^3}$
and repeating the simulation.
The resulting variation in the fitted attenuation response was taken as the systematic uncertainty due to the rock-density assumption.

\section{Data availability}
A list of selected events binned in 1-hour intervals,
together with the corresponding water-level data, is available at
\small {https://github.com/seojiwon0/Muography}.

\section{Code availability}
Tunnel scan CRY simulation codes are available at
\small {https://github.com/changhyon/hawlumusim}.

\section{Acknowledgments}
This work was supported by National Research Foundation of Korea~(NRF) grant funded
by the Korean government~(MSIT)(RS-2021-NR058750 and RS-2024-00438814).

\bibliographystyle{unsrt}

\begin{thebibliography}{10}

\bibitem{10.1098/rsta.1966.0024}
Walter~Heinrich Munk and David~Edgar Cartwright.
\newblock Tidal spectroscopy and prediction.
\newblock {\em Philosophical Transactions of the Royal Society of London,
  Series A: Mathematical and Physical Sciences}, 259(1105):533--581, 05 1966.

\bibitem{Morishima:2017ghw}
Kunihiro Morishima et~al.
\newblock {Discovery of a big void in Khufu's Pyramid by observation of
  cosmic-ray muons}.
\newblock {\em Nature}, 552(7685):386--390, 2017.

\bibitem{Tanaka:2023ami}
Hiroyuki K.~M. Tanaka et~al.
\newblock {Muography}.
\newblock {\em Nature Reviews Methods Primers}, 3(1):88, 2023.

\bibitem{Tanaka:2014}
Hiroyuki K.~M. Tanaka, Taro Kusagaya, and Hiroshi Shinohara.
\newblock {Radiographic visualization of magma dynamics in an erupting
  volcano}.
\newblock {\em Nature Communications}, 5(1):3381, 2014.

\bibitem{Tanaka2021}
Hiroyuki K.~M. Tanaka et~al.
\newblock First results of undersea muography with the tokyo-bay seafloor
  hyper-kilometric submarine deep detector.
\newblock {\em Scientific Reports}, 11(1):19485, Sep 2021.

\bibitem{Khaw:2025tor}
Kim~Siang Khaw et~al.
\newblock {Towards non-invasive sediment monitoring using muography: A pilot
  run at the Shanghai Outer Ring Tunnel}.
\newblock {\em Journal of Applied Physics}, 138(11):114902, 2025.

\bibitem{ParticleDataGroup:2024cfk}
S.~Navas et~al.
\newblock {Review of Particle Physics}.
\newblock {\em Physical Review D}, 110(3):030001, 2024.

\bibitem{10.1111/j.1365-246X.2010.04790.x}
N.~Lesparre et~al.
\newblock Geophysical muon imaging: feasibility and limits.
\newblock {\em Geophysical Journal International}, 183(3):1348--1361, 12 2010.

\bibitem{COSINE-100:2020jml}
H.~Prihtiadi et~al.
\newblock {Measurement of the cosmic muon annual and diurnal flux variation
  with the COSINE-100 detector}.
\newblock {\em Journal of Cosmology and Astroparticle Physics}, 02:013, 2021.

\bibitem{Abdullrahman2025}
Maghrabi Abdullrahman, Al~Harbi Hya, and Alruhaili Aied.
\newblock Exploring daily fluctuations of cosmic ray muon components at a low
  latitude site and their associations with space weather variables.
\newblock {\em Journal of Astrophysics and Astronomy}, 46(1):10, Jan 2025.

\bibitem{Seo:2024ijp}
J.~Seo et~al.
\newblock {Real-time portable muography with Hankuk Atmospheric-muon Wide
  Landscaping : HAWL}.
\newblock {\em Radiation Physics and Chemistry}, 225:112117, 2024.

\bibitem{Korea-Tunnel}
Young-ho Seo et~al.
\newblock {Introduction to the Design Case of Korea's First Undersea Road
  Tunnel (Boryeong Tunnel) - Focusing on Alignment Planning, Excavation, and
  Waterproofing Grouting}.
\newblock {\em Magazine of Korean Tunnelling and Underground Space
  Association(Korean)}, 12:30--38, 2010.

\bibitem{TideData2}
{Korea Hydrographic and Oceanographic Agency}.
\newblock Tidal observation data extreme highest tide level.
\newblock \url{https://khoa.go.kr/oceandata/oceaninfo/map.do}, 2026.
\newblock Boryeong tide station; accessed July 2026.

\bibitem{Korea-Tunnel2}
{Hyundai Engineering and Construction}.
\newblock [engineering drawing of the boryeong undersea tunnel].
\newblock Exhibit at the Boryeong Undersea Tunnel Museum.

\bibitem{Renault2022}
Lionel Renault and Patrick Marchesiello.
\newblock Ocean tides can drag the atmosphere and cause tidal winds over broad
  continental shelves.
\newblock {\em Communications Earth {\&} Environment}, 3(1):70, Mar 2022.

\bibitem{KVALE20065}
Erik~P. Kvale.
\newblock The origin of neap–spring tidal cycles.
\newblock {\em Marine Geology}, 235(1):5--18, 2006.
\newblock Proceedings of the 6th International Congress on Tidal Sedimentology
  (Tidalites 2004).

\bibitem{ART002186085}
Yong-Seok Seo, Hyun-Seok Yun, Donggyou Kim, and O-Il Kwon.
\newblock Analysis on physical and mechanical properties of rock mass in
  korea(korean).
\newblock {\em The journal of Engineering Geology}, 26(4):593--600, 2016.

\bibitem{Hagmann:2007ziw}
Chris Hagmann, David Lange, and Douglas Wright.
\newblock {Cosmic-ray shower generator (CRY) for Monte Carlo transport codes}.
\newblock {\em IEEE nuclear science symposium conference record}, 2:1143--1146,
  2007.

\bibitem{PughWoodworth2014}
David Pugh and Philip Woodworth.
\newblock {\em Sea-Level Science: Understanding Tides, Surges, Tsunamis and
  Mean Sea-Level Changes}.
\newblock Cambridge University Press, 2014.

\bibitem{ejps}
Eljen.
\newblock {\em Eljen Technology, GENERAL PURPOSE EJ-200, EJ-204, EJ-208,
  EJ-212}.
\newblock 2023.

\bibitem{sipm}
MPPC.
\newblock {\em MPPC for precision measurement, S13360-1375PE}.
\newblock 2024.

\bibitem{Be2004}
M.-M. B{\'e}, V.~Chist{\'e}, C.~Dulieu, E.~Browne, V.~Chechev, N.~Kuzmenko,
  R.~Helmer, A.~Nichols, E.~Sch{\"o}nfeld, and R.~Dersch.
\newblock {\em Table of Radionuclides}, volume~2 of {\em Monographie BIPM-5}.
\newblock Bureau International des Poids et Mesures, S{\`e}vres, 2004.
\newblock A = 151--242.

\bibitem{Knoll2010}
Glenn~F. Knoll.
\newblock {\em Radiation Detection and Measurement}.
\newblock John Wiley \& Sons, Hoboken, New Jersey, 4 edition, 2010.

\bibitem{TideData}
{Korea Hydrographic and Oceanographic Agency}.
\newblock Tidal observation data.
\newblock
  \url{https://www.khoa.go.kr/oceangrid/gis/category/reference/distribution.do},
  2025.

\bibitem{BAKER1984437}
Steve Baker and Robert~D. Cousins.
\newblock Clarification of the use of chi-square and likelihood functions in
  fits to histograms.
\newblock {\em Nuclear Instruments and Methods in Physics Research},
  221(2):437--442, 1984.

\bibitem{ERA5PressureLevels}
{Copernicus Climate Change Service}.
\newblock {ERA5} hourly data on pressure levels from 1940 to present, 2018.
\newblock Accessed February 2026.

\bibitem{Hersbach2020}
H.~Hersbach et~al.
\newblock The {ERA5} global reanalysis.
\newblock {\em Q. J. R. Meteorol. Soc.}, 146:1999--2049, 2020.

\bibitem{d3ea2d52-5ab2-3128-8b80-efb85267295d}
James~W. Cooley and John~W. Tukey.
\newblock An algorithm for the machine calculation of complex fourier series.
\newblock {\em Mathematics of Computation}, 19(90):297--301, 1965.

\bibitem{Harris1978}
Fredric~J. Harris.
\newblock On the use of windows for harmonic analysis with the discrete fourier
  transform.
\newblock {\em Proceedings of the IEEE}, 66(1):51--83, 1978.

\end{thebibliography}

\end{document}